\documentclass[pra, twocolumn, amsmath, amssymb, superscriptaddress]{revtex4-2}

\usepackage{graphicx, braket, xcolor}
\usepackage{amsmath}
\usepackage{amssymb}
\usepackage{comment}
\usepackage[pdftex,colorlinks=true,citecolor=black,urlcolor=black,linkcolor=black]{hyperref}
\usepackage{ulem}
\usepackage[inkscapelatex=false]{svg}

\usepackage[mathscr]{euscript}

\begin{document}
\title{Controlling the dynamical phase diagram of a spinor BEC using time-dependent potentials}

\author{Q. Guan}
\affiliation{Department of Physics and Astronomy, Washington State University, Pullman, WA 99164, USA}
\author{D. Blume}
\affiliation{Homer L. Dodge Department of Physics and Astronomy, The University of Oklahoma, Norman, Oklahoma 73019, USA}
\affiliation{Center for Quantum Research and Technology, The University of Oklahoma, Norman, Oklahoma 73019, USA}
\author{R.~J. Lewis-Swan}
\affiliation{Homer L. Dodge Department of Physics and Astronomy, The University of Oklahoma, Norman, Oklahoma 73019, USA}
\affiliation{Center for Quantum Research and Technology, The University of Oklahoma, Norman, Oklahoma 73019, USA}

\begin{abstract}
    We theoretically investigate the spin-mixing dynamics of a spinor BEC subject to a time-dependent confining potential. Our study provides a theory framework for the experimental results reported in Phys. Rev. A {\bf 109}, 043309 (2024). We exploit the disparity of energy scales associated with the spatial and internal (spin) degrees of freedom under typical experimental conditions to develop an effective few-mode description of the spin dynamics. Our model demonstrates how the details of the potential, such as driving frequency and amplitude, can be used to independently control spin-changing and spin-preserving collision processes as well as the effective Zeeman energy of the internal states. We obtain the dynamical phase diagram of the effective model and discuss how its structure is altered relative to a spinor BEC with frozen spatial degrees of freedom. The applicability of our effective model is verified through Gross-Pitaevskii simulations that capture the interplay of spin and spatial degrees of freedom, and we identify parameter regimes that can be feasibly explored by future experiments. Our findings highlight the utility of dynamical confining potentials for the control of non-equilibrium spin-mixing dynamics in spinor BECs. 
\end{abstract}

\maketitle 

\section{Indroduction}
Spinor Bose gases provide a highly controllable, well-isolated platform for quantum simulation of a wide variety of non-equilibrium phenomena \cite{StamperKurn2013spinor,Polkovnikov2011nonequilibrium,kawaguchi_2012}. A large portion of experimental and theoretical investigations specifically focus on spin-mixing dynamics, where atoms undergo coherent interconversion between different Zeeman states driven by two-body collisions. These have been used to shed light on nonlinear dynamics \cite{Zhang2005,Evrard2019shapiro,Li2019cavityspinor,Zhang2023spinorbit,Fujimoto2019floquetspinor}, ground and excited-state quantum phase transitions \cite{Jiang2014adiabatic,Klempt2023eqpt}, related dynamical phase transitions \cite{Duan_DPT_2020,Duan_DPT_2019,Dag_2018}, and thermalization or lack thereof \cite{Dag_2024_scars,AustinHarris2024scars}. In addition, spin-mixing has been explored for potential quantum science applications such as many-body state preparation \cite{Gross2011a,Oberthaler2022entanglement,Klempt2015epr} and quantum-enhanced sensing \cite{Lucke_2011,Linnemann_2016,You2018spin1dicke,Hamley2012a,Mao2023-xe,Anders_2021_momentum}. 
Studies of spin-mixing dynamics in these works have leveraged the large disparity in energy scales between the spin and spatial degrees of freedom. The spatial dynamics are typically assumed to be frozen out of the problem, leaving the dynamics of the spin degree of freedom to be dictated by the interplay of two parameters: The effective spin-spin interaction coefficient $c_2$ and the single-particle quadratic Zeeman energy $q$ \cite{Law_1998,Zhang2005,Liu2014}. 

In recent work by us and collaborators~\cite{Hardesty2023,Austin2024}, it was theoretically and experimentally demonstrated that the assumption of frozen spatial dynamics could be loosened while retaining an effective spin-only description. A time-dependent spatial density profile was shown to present additional opportunities for control over the spin-mixing dynamics. Here, we introduce a detailed theoretical framework demonstrating how periodically modulated trapping potentials can be used to expand the accessible dynamical phase diagram of spin-mixing in spinor gases. In particular, we demonstrate that the relative strength of distinct spin-preserving and spin-changing collision processes can be controlled through the amplitude of the potential modulation, while the quadratic Zeeman energy is tuned via the driving frequency. We develop an effective model that describes a series of dynamical phase transitions at absolute Zeeman energies defined by half-integer multiples of the driving frequency. We show that these dynamical phase transitions are generalizations of those previously studied in spinor BECs \cite{Zhang2005,Duan_DPT_2020,Duan_DPT_2019,Dag_2018}. To complement our effective model, which is based on a simplified few-mode description after integrating out the independent spatial dynamics, we present results of more sophisticated Gross-Pitaevskii simulations that capture the dynamical interplay of spin and spatial degrees of freedom. We use these simulations to systematically investigate the validity of our effective few-mode model and identify parameter regimes that can be explored in state-of-the-art spinor BEC experiments.

Our work complements prior investigations or proposals for Floquet engineered dynamics in spinor BECs using alternative controls, such as applied microwave or magnetic fields \cite{Evrard2019shapiro,Fujimoto2019floquetspinor,Zhang2023spinorbit} and atom-light interactions \cite{Li2019cavityspinor}. From a fundamental perspective, independent control of the different spin-spin interactions expands the dynamical phase diagram along a new dimension in parameter space and opens new pathways for the generation of entanglement, and can unlock new approaches for the exploitation of spinor gases for quantum sensing. In particular, our proposal to effectively shift the Zeeman energy via modulated trapping potentials allows the study of spin-mixing dynamics while working at large magnetic fields in a distinct fashion to prior work, which used microwave dressing and is constrained by heating and atom loss \cite{Liu2014}. Collectively, these features can lead to new approaches for the exploitation of spinor gases for quantum sensing.

The remainder of the paper is outlined as follows. Section~\ref{sec:model} introduces the basic theoretical model and an effective few-mode description of a spinor BEC in a periodically modulated potential. We then systematically map out the spin-mixing dynamical phase diagram in Sec.~\ref{sec:DynamicalPhaseDiagram} using a combination of analytic and numerical calculations. Section \ref{sec:GPE} provides a detailed benchmarking of our effective model and predictions through Gross-Pitaevskii simulations that capture the interplay of spin and spatial degrees of freedom.  We also present a discussion of the parameter regimes that can be accessible in current spinor BEC experiments. In Sec.~\ref{sec:Discussion}, we conclude with a discussion of our results in the context of other recent works that have studied spinor BECs with driving fields and comment on future prospects.

\section{Theoretical description of spinor BEC in time-dependent potentials \label{sec:model}}
In this section we introduce the theoretical description of the system under investigation -- a condensate of $N$ spin-$1$ atoms confined within a trapping potential $V(\mathbf{r},t)$ that is periodically modulated in time. The disparate energy scales of the spatial and spin degrees of freedom enable us to construct a simplified description of the nonlinear spin-mixing dynamics of the condensate, wherein the modulation of the trapping potential generates a time-dependent spin-spin interaction. We identify that the time-dependent interaction leads to a generalized model of spin-mixing, which possesses parameters that can be tuned by varying the properties of the modulated potential.  


\subsection{Multi-mode description of spin and spatial dynamics}
The complete Hamiltonian describing the spin and spatial degrees of freedom can be decomposed into spin-independent and spin-dependent pieces \cite{kawaguchi_2012,StamperKurn2013spinor},
\begin{equation}\label{eqn:FullHam}
    \hat{H} = \hat{H}_0 + \hat{H}_{\mathrm{spin}} .
\end{equation}
The first term includes contributions from the kinetic energy, an arbitrary trapping potential $V(\mathbf{r},t)$ and spin-independent density-density interactions, 
\begin{widetext}
\begin{equation}\label{eqn:SpatialHam}
    \hat{H}_0 = \int d^3\mathbf{r} \sum_{m=-1,0,1} \hat{\psi}^{\dagger}_m(\mathbf{r})\left[ -\frac{\hbar^2}{2M}\nabla^2 + V(\mathbf{r},t)\right] \hat{\psi}_m(\mathbf{r}) + \frac{g_0}{2}\int d^3\mathbf{r} \sum_{m,m^{\prime} = -1,0,1} \hat{\psi}^{\dagger}_m(\mathbf{r}) \hat{\psi}^{\dagger}_{m^{\prime}}(\mathbf{r}) \hat{\psi}_{m^{\prime}}(\mathbf{r}) \hat{\psi}_m(\mathbf{r}) .
\end{equation}
\end{widetext}
Here, $\hat{\psi}_m$ is the atomic field annihilation operator associated with the $m = 0,\pm1$ Zeeman sublevels, $g_0$ is the spin-independent interaction coefficient and $M$ is the atomic mass. 
The second term on the RHS of Eq.~(\ref{eqn:FullHam}),
\begin{widetext}
\begin{multline}\label{eqn:SpinHam}
    \hat{H}_{\mathrm{spin}} = \int d^3\mathbf{r} \left\{~\frac{g_2}{2} \left[ 2\hat{\psi}^{\dagger}_0(\mathbf{r})\hat{\psi}^{\dagger}_0(\mathbf{r})\hat{\psi}_1(\mathbf{r})\hat{\psi}_{-1}(\mathbf{r}) + 
    2\hat{\psi}^{\dagger}_1(\mathbf{r})\hat{\psi}^{\dagger}_{-1}(\mathbf{r})\hat{\psi}_0(\mathbf{r})\hat{\psi}_{0}(\mathbf{r}) + 2\hat{\psi}^{\dagger}_0(\mathbf{r})\hat{\psi}^{\dagger}_1(\mathbf{r})\hat{\psi}_1(\mathbf{r})\hat{\psi}_0(\mathbf{r}) \right. \right.  \\
    + 2\hat{\psi}^{\dagger}_0(\mathbf{r})\hat{\psi}^{\dagger}_{-1}(\mathbf{r})\hat{\psi}_{-1}(\mathbf{r})\hat{\psi}_0(\mathbf{r}) 
    + \hat{\psi}^{\dagger}_1(\mathbf{r})\hat{\psi}^{\dagger}_1(\mathbf{r})\hat{\psi}_1(\mathbf{r})\hat{\psi}_1(\mathbf{r}) + \hat{\psi}^{\dagger}_{-1}(\mathbf{r})\hat{\psi}^{\dagger}_{-1}(\mathbf{r})\hat{\psi}_{-1}(\mathbf{r})\hat{\psi}_{-1}(\mathbf{r})
    \\
    \left. \left.  - 2\hat{\psi}^{\dagger}_{1}(\mathbf{r})\hat{\psi}^{\dagger}_{-1}(\mathbf{r})\hat{\psi}_{-1}(\mathbf{r})\hat{\psi}_{1}(\mathbf{r}) \right] + q\left[ \hat{\psi}^{\dagger}_1(\mathbf{r})\hat{\psi}_1(\mathbf{r}) + \hat{\psi}^{\dagger}_{-1}(\mathbf{r})\hat{\psi}_{-1}(\mathbf{r}) \right] \right\} ,
\end{multline}
\end{widetext}
accounts for spin-dependent interactions with strength $g_2$ and the quadratic Zeeman energy $q$.

This theoretical model of the spinor BEC is simplified by exploiting that the energy scales associated with $\hat{H}_0$, i.e., the trapping potential and spin-independent interactions, are typically much larger than those of the spin-mixing Hamiltonian $\hat{H}_{\mathrm{spin}}$. 
The separation of energy scales motivates a single spatial-mode approximation (SMA), in which the bosonic field operators are replaced by $\hat{\psi}_m(\mathbf{r},t) = \hat{a}_m \phi(\mathbf{r},t)$, where $\hat{a}_m$ is the bosonic annihilation operator for the Zeeman state $m$ associated with a common, spin-independent spatial mode described by $\phi(\mathbf{r},t)$ \cite{Law_1998}. In this approximation, the spatial dynamics are captured by $\phi(\mathbf{r},t)$, which is assumed to be described by the time-dependent scalar Gross-Pitaevskii equation obtained from $\hat{H}_0$. Separately, the spin-mixing dynamics are obtained by substituting the SMA decomposition of the field operators into $\hat{H}_{\mathrm{spin}}$ and integrating out the spatial co-ordinates to yield the effective few-mode Hamiltonian, 
\begin{widetext}
\begin{equation}\label{eqn:H3mode}
    \hat{H}_{\mathrm{eff}}(t) = \frac{c_2(t)}{N} \left( \hat{a}_0^{\dagger}\hat{a}_0^{\dagger}\hat{a}_1\hat{a}_{-1} + h.c. \right) + \frac{c_2(t)}{N} \hat{a}^{\dagger}_0\hat{a}_0 \left( \hat{a}^{\dagger}_1\hat{a}_1 + \hat{a}^{\dagger}_{-1}\hat{a}_{-1} \right) + \frac{c_2(t)}{2N}\left( \hat{a}^{\dagger}_1\hat{a}_1 - \hat{a}^{\dagger}_{-1}\hat{a}_{-1} \right)^2 
    + q \left( \hat{a}^{\dagger}_1\hat{a}_1 + \hat{a}^{\dagger}_{-1}\hat{a}_{-1} \right) , 
\end{equation}
\end{widetext}
with the (time-dependent) spin-dependent interaction coefficient, 
\begin{equation}\label{eqn:c2_t}
    c_2(t) = g_2 N \int d^3\mathbf{r} \vert \phi(\mathbf{r},t) \vert^4 .
\end{equation}
Inspection of the above equation shows that time evolution of $c_2(t)$ arises from the dynamics of the spatial density profile $\rho(\mathbf{r},t) = \vert \phi(\mathbf{r},t) \vert^2$. As a result, Eqs.~(\ref{eqn:H3mode}) and (\ref{eqn:c2_t}) are collectively referred to as a description of spin-mixing under a \textit{dynamical} single spatial-mode (dSMA) approximation \cite{Hardesty2023}.

\subsection{Effective few-mode model of spin-mixing}
The time-dependence of $c_2(t)$, and thus $\rho(\mathbf{r},t)$, is crucial for understanding the spin-mixing dynamics generated by the few-mode Hamiltonian (\ref{eqn:H3mode}). To proceed we thus introduce a specific parametrization of the time-dependent potential $V(\mathbf{r},t)$.

\begin{figure}[t]
    \centering
    \includegraphics[width=8cm]{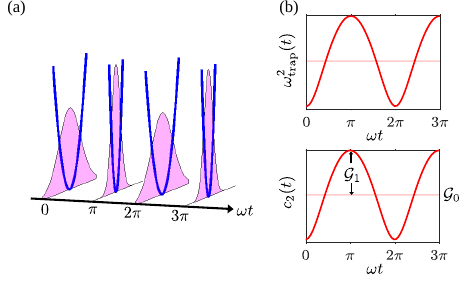}
    \caption{(a) Example schematic for the proposed driving protocol. Slowly compressing and relaxing a harmonic potential $V_0(\mathbf{r},t)$ with period $T = 2\pi/\omega$ (blue line, shown as a $1$D cut along an arbitrary axis) leads to a commensurate modulation of the density profile of the confined spinor BEC (indicated by magenta shaded region). (b) Example timetrace of the modulated trapping frequency, $\omega^2_{\mathrm{trap}} = \omega^2_{\mathrm{osc},0} + \omega^2_{\mathrm{osc},1}\cos(\omega t - \varphi)$ with $\varphi = \pi$ (top panel). Within the assumptions of our model, this generates a time-dependent spin-spin interaction strength $c_2(t)$ (bottom panel), which may be expanded as a Fourier series as in Eq.~(\ref{eqn:c2_ansatz}). In this case, the time-averaged value of $c_2(t)$ gives rise to the effective interaction strength $\mathcal{G}_0$, while the amplitude of the variation is related to $\mathcal{G}_1$. (For this example $\mathcal{G}_{k>1} = 0$.) The illustrations in (b) use arbitrary units for $\omega^2_{\mathrm{trap}}$ and $c_2(t)$.
    }
    \label{fig:Setup}
\end{figure}

We assume that the trapping potential for the condensate is periodically varied so that $V(\mathbf{r},t+T) = V(\mathbf{r},t)$, where $T$ is the driving period. Furthermore, we assume that the driving period is: i) \textit{long} compared to the inverse energy scales of $\hat{H}_0$ and ii) \textit{short} compared to the typical timescales of $\hat{H}_{\mathrm{eff}}(t)$. The separation of energy scales between the spin and spatial degrees of freedom, combined with accessible dynamical control of optical trapping potentials, enables these conditions to be simultaneously met in experimentally realistic settings. The importance of condition i) is that it enables us to assume that the evolution of the spatial density profile follows the potential approximately adiabatically, which means that the density profile evolves periodically, $\rho(\mathbf{r},t+T) = \rho(\mathbf{r},t)$. In turn, Eq.~(\ref{eqn:c2_t}) then dictates that the time-dependent spin-spin interaction coefficient will also vary periodically, $c_2(t+T) = c_2(t)$. 

Possible scenarios to realize the modulated potential include using a standing wave optical lattice with time-dependent amplitude or a harmonic trap with time-dependent frequency. In the former, a spinor condensate is loaded into a one-dimensional standing wave lattice of the form $V(\mathbf{r},t) = V_0(t) \cos^2(\mathbf{k}_L \cdot \mathbf{r})$, where $\mathbf{k}_L$ is the lattice wavevector and the lattice amplitude is modulated sinusoidally according to 
\begin{equation}
    V_0(t) = V_{\mathrm{osc},0} + V_{\mathrm{osc},1}\cos(\omega t - \varphi) ,
\end{equation}
where $V_{\mathrm{osc},0}$ and $V_{\mathrm{osc},1}$ control the minimum/maximum amplitude of the lattice, $\varphi$ is a tunable phase offset and $\omega = 2\pi/T$ is the driving frequency. This scenario was realized experimentally in Ref.~\cite{Austin2024} for the case $V_{\mathrm{osc},0} = V_{\mathrm{osc},1}$.
Alternatively, the condensate is prepared in a harmonic trap (see Fig.~\ref{fig:Setup}) characterized by the potential 
\begin{eqnarray}
\label{trap_potential}
    V(\mathbf{r},t) = \frac{1}{2}M\omega^2_{\mathrm{trap}}(t)\mathbf{r}^2,
\end{eqnarray}
where
\begin{equation}
\label{trap_freq}
    \omega^2_{\mathrm{trap}}(t) = \omega^2_{\mathrm{osc},0} + \omega^2_{\mathrm{osc},1}\cos(\omega t - \varphi)
\end{equation}
is the time-dependent trapping frequency, with the constraint that the control parameters satisfy $\omega^2_{\mathrm{osc},0}\geq\omega^2_{\mathrm{osc},1}$. As a result of the modulated confining potential, the spatial density of the condensate changes over time such that $c_2(t)$ varies periodically between minimum and maximum values, as illustrated in Fig.~\ref{fig:Setup}(b) for the harmonic trap. 

We choose to enforce an isotropic harmonic trap and driving for simplicity of illustration, and such a scenario has been demonstrated experimentally in, e.g., Ref.~\cite{Lobser2015breathing}. It would be straightforward to consider extensions to an anisotropic trap or modulation along a subset of only one or two trap axes within the framework of our effective few-mode model.




To make the above discussion precise, we express the time-dependence of the spin-dependent interaction as a Fourier series, 
\begin{equation}\label{eqn:c2_ansatz}
    c_2(t) = \mathcal{G}_0 + \sum_{k=1}^{\infty} \mathcal{G}_k \cos \left( k\omega t - \varphi_k \right) .
\end{equation}
Without loss of generality we take $\mathcal{G}_0,\mathcal{G}_k \geq 0$ by appropriate definition of the phases $\varphi_k$~\footnote{This ensures $c_2(t)$ is strictly positive in our case. In general, Eq.~(\ref{eqn:c2_t}) enforces only that $c_2(t)$ is of fixed sign, but predictions for $c_2(t) < 0$ ($g_2 < 0$) can be obtained by setting $q \to -q$ throughout our calculations.}. Substitution of the expansion (\ref{eqn:c2_ansatz}) into Eq.~(\ref{eqn:H3mode}) and transforming into the interaction picture with respect to the Zeeman energy $q$ yields: 
\begin{widetext}
\begin{multline}\label{eqn:DrivenH3mode}
    \hat{H}_I(t) = \frac{\mathcal{G}_0}{N}\left[ e^{-2iqt/\hbar}\hat{a}_0^{\dagger}\hat{a}_0^{\dagger}\hat{a}_1\hat{a}_{-1} + e^{2iqt/\hbar}\hat{a}^{\dagger}_1\hat{a}^{\dagger}_{-1}\hat{a}_0\hat{a}_0 + \hat{a}^{\dagger}_0\hat{a}_0\left( \hat{a}^{\dagger}_1\hat{a}_1 + \hat{a}^{\dagger}_{-1}\hat{a}_{-1} \right) + \left( \hat{a}^{\dagger}_1\hat{a}_1 -\hat{a}^{\dagger}_{-1}\hat{a}_{-1} \right)^2 \right]  \\
    + \sum_{k=1}^{\infty} \frac{\mathcal{G}_k}{2N}\Bigg[ \left(e^{-i(k\omega+2q/\hbar)t + i\varphi_k} + e^{i(k\omega - 2q/\hbar)t - i\varphi_k} \right)\hat{a}_0^{\dagger}\hat{a}_0^{\dagger}\hat{a}_1\hat{a}_{-1} + \left(e^{i(k\omega+2q/\hbar)t - i\varphi_k} + e^{-i(k\omega - 2q/\hbar)t + i\varphi_k} \right)\hat{a}^{\dagger}_1\hat{a}^{\dagger}_{-1}\hat{a}_0\hat{a}_0 \\
    + \left( e^{-i(k\omega t-\varphi_k)} + e^{i(k\omega t - \varphi_k)} \right) \hat{a}^{\dagger}_0\hat{a}_0\left( \hat{a}^{\dagger}_1\hat{a}_1 + \hat{a}^{\dagger}_{-1}\hat{a}_{-1} \right) + \frac{1}{2}\left( e^{-i(k\omega t-\varphi_k)} + e^{i(k\omega t - \varphi_k)} \right) \left( \hat{a}^{\dagger}_1\hat{a}_1 - \hat{a}^{\dagger}_{-1}\hat{a}_{-1} \right)^2 \Bigg] .
\end{multline}
\end{widetext}

The form of $\hat{H}_I(t)$ can be greatly simplified when the Zeeman energy is tuned near one of a series of resonances given by half-integer multiples of the driving frequency, i.e., $q/\hbar  \approx j\omega/2$ with $j \in \mathbb{Z}$. For $j=0$ and $\vert q \vert \gg \mathcal{G}_k$, we may make a rotating-wave approximation and neglect the contributions from the second and third lines of Eq.~(\ref{eqn:DrivenH3mode}). In this case the remaining Hamiltonian reduces back to Eq.~(\ref{eqn:H3mode}) with the identification $c_2(t) \to \mathcal{G}_0$ and is equivalent to the standard SMA description obtained from a static condensate in a time-independent potential. For $j \neq 0$, retaining only the near-resonant oscillatory terms after a rotating-wave approximation and making a unitary transformation $\hat{H}_I(t) \to \hat{U}_j \hat{H}_I(t) \hat{U}^{\dagger}_j - i\hbar\hat{U}_j^{\dagger}\partial_t\hat{U}_j$ with $\hat{U}_j = e^{i(q/\hbar -j\omega/2)t(\hat{a}^{\dagger}_1\hat{a}_1 + \hat{a}^{\dagger}_{-1}\hat{a}_{-1})}$ leads to the Hamiltonian, 
\begin{multline}\label{eqn:Heff}
    \hat{H}_{\mathrm{eff},j} = \frac{\mathcal{G}_0}{N} \hat{a}^{\dagger}_0\hat{a}_0\left( \hat{a}^{\dagger}_1\hat{a}_1 + \hat{a}^{\dagger}_{-1}\hat{a}_{-1} \right) + \frac{\mathcal{G}_0}{2N}\left( \hat{a}^{\dagger}_1\hat{a}_1 - \hat{a}^{\dagger}_{-1}\hat{a}_{-1} \right)^2  \\
    + \frac{\mathcal{G}_j}{2N}\left( e^{-i\varphi_j}\hat{a}^{\dagger}_0 \hat{a}^{\dagger}_0 \hat{a}_1 \hat{a}_{-1} + e^{i\varphi_j}\hat{a}_0 \hat{a}_0 \hat{a}^{\dagger}_1 \hat{a}^{\dagger}_{-1} \right) \\
    + q_{\mathrm{eff},j} \left( \hat{a}^{\dagger}_1\hat{a}_1 + \hat{a}^{\dagger}_{-1}\hat{a}_{-1} \right) , 
\end{multline}
where $q_{\mathrm{eff},j} = q - j\hbar\omega/2$ is an effective Zeeman energy. The magnetization $\hat{\mathcal{M}} = \hat{a}^{\dagger}_1\hat{a}_1 - \hat{a}^{\dagger}_{-1}\hat{a}_{-1}$ commutes with $\hat{H}_{\mathrm{eff},j}$ and is a conserved quantity. In the remainder of the paper we restrict our focus to the zero magnetization sector for simplicity. 

Equation (\ref{eqn:Heff}) describes a model of spin-mixing with: i) tunable relative contributions from spin-preserving (first line) and spin-changing (second line) collision processes, ii) tunable spinor phase $\varphi_j$ controlled through the modulation of the potential, and iii) an effective Zeeman energy $q_{\mathrm{eff},j}$ that is shifted by the modulation frequency of the applied potential. The first feature leads to a richer dynamical phase diagram of spin-mixing dynamics, with potential implications for studies of, e.g., dynamical phase transitions -- which we discuss in the following Sec.~\ref{sec:DynamicalPhaseDiagram} -- and preparation of entangled states \cite{SpinorQFI}. Related to this, indirect control of the spinor phase through the potential, as opposed to directly imprinting an initial relative phase between the Zeeman components during, e.g., state preparation, provides an additional tool for control of spin-mixing dynamics. Lastly, the ability to shift the Zeeman energy in a controlled manner over comparatively large scales (bounded from above by the energy scale of the spatial dynamics) can enable the study of spin-mixing dynamics at larger field strengths than otherwise accessible, and complements other techniques that have been exploited to shift the Zeeman energy such as microwave dressing \cite{Bloch2006,Liu2014}.



\section{Classification of dynamical phase diagram\label{sec:DynamicalPhaseDiagram}}
Nonlinear spin-mixing is known to lead to a variety of non-equilibrium phenomenona including dynamical phase transitions (DPTs) \cite{Dag_2018,Duan_DPT_2019,Duan_DPT_2020}. Analogous to equilibrium phase transitions, DPTs are characterized by a time-averaged order parameter that distinguishes phases of different dynamical behavior and features nonanalytic behavior at the critical point \cite{Jamir2022}. In this section, we map out the dynamical phase diagram of Eq.~(\ref{eqn:Heff}) in the classical limit and identify how the tunable relative strengths of spin-preserving and spin-changing collisions influences and alters the dynamics of the system, as compared to the standard SMA description of spin-mixing. Our analysis of the dynamics in the classical limit is well motivated, as many spinor BEC experiments work with large condensates and investigate quench dynamics initiated from states where quantum fluctuations can be neglected to a first approximation. We will investigate the relevance of quantum fluctuations in the context of Eq.~(\ref{eqn:Heff}) in a future manuscript \cite{SpinorQFI}, but note that there has been extensive recent work to develop a more rigorous quantum framework of DPTs, including their relation to quantum phase transitions in excited states \cite{Qingze2021,Relano2022,Relano2023,RLS_Dicke_2021}.

\subsection{Classical dynamics and phase portrait}
We focus on quench dynamics initiated from a spin coherent state, in which a fraction of atoms from an initially pure condensate of $m = 0$ atoms are coherently coupled by an RF or microwave pulse to the $m=\pm 1 $ states \cite{Jianwen_2019,Wrubel_2018}. The quench dynamics from such initial states are well described by the evolution of two classical variables: the fractional population of the $m = 0$ mode, $\rho_0 = \langle \hat{a}^{\dagger}_0 \hat{a}_0 \rangle/N$, and the spinor phase, $\theta = \mathrm{Arg}\left(\hat{a}_0\hat{a}_0\hat{a}^{\dagger}_1\hat{a}^\dagger_{-1} \right)$. 
In terms of these variables, the classical limit of Eq.~(\ref{eqn:Heff}) for a given resonance with $q/\hbar \approx j\omega/2$ is governed by the Hamiltonian~\footnote{In more detail, we replace operators by c-numbers in Eq.~(\ref{eqn:Heff}), $\hat{a}_m \to \alpha_m$, eliminate conserved quantities and then make a variable change with $\rho_0 = \vert\alpha_0\vert^2/N$ and $\theta = \mathrm{Arg}\left(\alpha_0^2 \alpha_1^*\alpha_{-1}^* \right)$.}, 
\begin{equation}\label{eqn:Hmf}
    \frac{H_{\mathrm{mf},j}}{\mathcal{G}_0} = \rho_0(1-\rho_0) \left[ 1 + \frac{\eta}{2} \cos (\theta + \varphi_j) \right] + \tilde{q}( 1 - \rho_0 ) , 
\end{equation}
with associated equations of motion, 
\begin{equation}\label{eqn:classicalEOM}
    \begin{gathered}
            \frac{\hbar}{\mathcal{G}_0}\dot{\rho_0} = \eta(1-\rho_0)\rho_0\sin(\theta + \varphi_j) , \\
            \frac{\hbar}{\mathcal{G}_0}\dot{\theta} = -2\tilde{q} + (1-2\rho_0)\left[2 + \eta\cos(\theta+\varphi_j)\right] .
    \end{gathered}
\end{equation}
The dynamics of the classical model depend on two key parameters: $\eta = \mathcal{G}_j/\mathcal{G}_0$ and $\tilde{q} = q_{\mathrm{eff},j}/\mathcal{G}_0$. We note that within the context of the physical system outlined in Sec.~\ref{sec:model}, the ratio of the interaction strengths is physically constrained to $\eta \leq 1$ to ensure $ c_2(t)$ has a fixed sign, consistent with the definition of Eq.~(\ref{eqn:c2_t}).  
We highlight that the form of both Eqs.~(\ref{eqn:Hmf}) and (\ref{eqn:classicalEOM}) illustrate how the phase $\varphi_j$ that is introduced by the driving can be used to control the initial spinor phase, specifically by absorbing it into the definition of the spinor phase as an offset $\theta \to \theta + \varphi_j$.

\begin{figure}[tb]
    \centering
    \includegraphics[width=8cm]{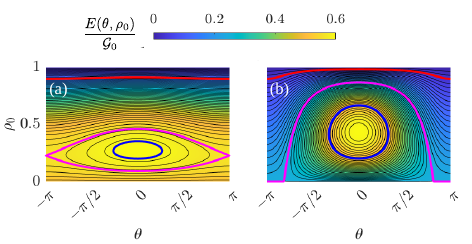}
    \caption{Example phase portraits of the spin-mixing dynamics for $\tilde{q} = 0.49$ and (a) $\eta = 0.2$ ($\tilde{q} < 1 - \eta/2$) and (b) $\eta = 1.8$ ($\tilde{q} > 1 - \eta/2$). We choose $\varphi_j = 0$ in both panels. Background heat map indicates the energy defined by Eq.~(\ref{eqn:Hmf}). We emphasize typical trajectories for the ID and ZD phases using initial conditions $\rho_0 = 0.2$ (blue lines) and $\rho_0 = 0.95$ (red lines) in both panels. Separatrices (magenta lines) are defined by the critical energy (a) $E_c/\mathcal{G}_0 = (2+2\tilde{q}-\eta)/[8(2-\eta)]$ and (b) $E_c/\mathcal{G}_0 = \tilde{q}$, which are obtained using the associated fixed points defined in Eqs.~(\ref{eqn:SaddlePoints_smalleta}) and (\ref{eqn:SaddlePoints_largeeta1}).} 
    \label{fig:PhasePortrait}
\end{figure}

Example phase portraits generated from Eqs.~(\ref{eqn:Hmf}) and (\ref{eqn:classicalEOM}) are shown in Fig.~\ref{fig:PhasePortrait} for $\tilde{q} = 0.49$ and two different values of the relative interaction strength: panel (a) $\eta = 0.2$ and panel (b) $\eta = 1.8$. The trajectories in phase space are delineated into two regimes split by a separatrix, analogous to a non-rigid pendulum, and can be classified according to the behavior of the spinor phase $\theta$: i) a \textit{Zeeman-dominated} (ZD) regime where $\theta$ grows unbounded and ii) an \textit{Interaction-dominated} (ID) regime where the spinor phase is bounded \cite{Zhang2005}. Example trajectories for each class are highlighted by the red (ZD) and blue (ID) lines in each panel of Fig.~\ref{fig:PhasePortrait}. 

We note that while our choice of $\eta = 1.8$ in panel (b) is unphysical within our framework (i.e., it violates the condition that $c_2$ has a fixed sign), we include it nevertheless. 
This is motivated by recent work in Ref.~\cite{Fujimoto2019floquetspinor} that studied the ground-state phase diagram of a spinor BEC subject to a fast oscillating Zeeman shift $q(t)$ realized through microwave dressing (see also Sec.~\ref{sec:Discussion}). In that case, Eq.~(\ref{eqn:Heff}) similarly emerges as an effective description but one is constrained to the regime $2 - \eta \ll 1$ (i.e., $\mathcal{G}_1 \approx 2\mathcal{G}_0$ in our framework). Given this context, it is thus of general interest to investigate the dynamics of the effective Hamiltonian (\ref{eqn:Hmf}) beyond the immediate scope of what may be achievable only through time-dependent confining potentials.

\begin{figure*}[tb]
    \centering
    \includegraphics[width=16cm]{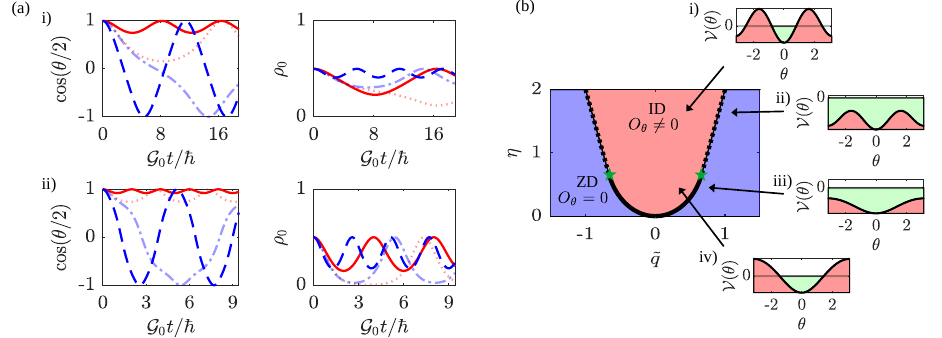}
    \caption{(a) Example time-traces for the conjugate variables $(\rho_0,\theta)$ for i) $\eta = 0.2$ and ii) $\eta = 1.8$ (same as phase portraits in Fig.~\ref{fig:PhasePortrait}) starting from the initial state with $\rho_0(0) = 1/2$ and $\theta(0)+\varphi_j = 0$. We compare dynamics for $\tilde{q} = 0.7\tilde{q}_{c,+}$ (red solid lines), $0.99\tilde{q}_{c,+}$ (faded red dotted lines), $1.02\tilde{q}_{c,+}$ (faded blue dot-dashed lines) and $1.5\tilde{q}_{c,+}$ (dashed blue lines), where $\tilde{q}_{c,+}$ is the critical Zeeman energy [see Eqs.~(\ref{eqn:qc_case1}) and (\ref{eqn:qc_case2}) and surrounding discussion in text]. 
    (b) Dynamical phase diagram, characterized by $O_{\theta} = \overline{\cos(\theta/2)}$ (see text), as a function of relative interaction strength $\eta$ and normalized Zeeman shift $\tilde{q}$, obtained for the initial condition $\rho_0(0)=1/2$ and $\theta(0) + \varphi_j = 0$. ID and ZD phases are indicated by red and blue shaded regions, respectively (see labels). The DPT separating these phases is indicated by the solid and dotted black lines, which correspond to $\tilde{q}_{c,\pm}$ for $\eta < \eta_{c,\pm}$ [Eq.~(\ref{eqn:qc_case1})] and $\eta > \eta_{c,\pm}$ [Eq.~(\ref{eqn:qc_case2})], respectively. For the chosen initial condition, $\tilde{q}_{c,+}$ ($\tilde{q}_{c,-}$) delineates the ID and ZD phases for positive (negative) normalized Zeeman shift. We additionally indicate the values of $\eta_{c,+} = \eta_{c,-} = 2/3$ on the phase boundaries with green stars.  
    Call-outs i)-iv) show representative $1$D potentials $\mathcal{V}(\theta)$ (plotted in units of $\mathcal{G}_0$, see Appendix~\ref{app:potentials}) for the indicated parameter regimes. The green and red shaded regions indicate the accessible and inaccessible regions of the spinor phase $\theta$, respectively. Faded horizontal line in each plot indicates $\mathcal{V}(\theta) = 0$ to highlight the turning points of the dynamics. 
    } 
    \label{fig:DynamicalPhaseDiagram}
\end{figure*}

The relative strength of the spin-changing and spin-preserving interactions, characterized by $\eta$, and the Zeeman shift $\tilde{q}$ influence the structure of the phase portrait and the location of the separatrix, indicated by the magenta lines in Fig.~\ref{fig:PhasePortrait}. Restricting to $\varphi_j = 0$ and $\eta < 2$ for simplicity, it is straightforward to show that for $\vert \tilde{q} \vert \leq 1 - \eta/2$ the separatrix 
passes through fixed points located at 
\begin{equation}\label{eqn:SaddlePoints_smalleta}
    (\rho_0, \theta)_{*} = \left(\frac{1}{2} - \frac{\tilde{q}}{2-\eta}, [2n+1]\pi \right) , 
\end{equation}
with $n \in \mathbb{Z}$.
When $1 - \eta/2 < \vert \tilde{q} \vert \leq 1 + \eta/2$, the separatrix instead passes through the extremal values of $\rho_0$ according to the sign of the effective Zeeman shift. For $\tilde{q} > 0$, the fixed points are
\begin{equation}\label{eqn:SaddlePoints_largeeta1}
    (\rho_0, \theta)_{*} = 
        \left(0, \pm\mathrm{atan}\left[ \frac{\sqrt{\eta^2 - 4(1-\tilde{q})^2}}{2(1-\tilde{q})} \right] + 2n\pi \right) ,
\end{equation}
while for $\tilde{q} < 0$, they are given by
\begin{equation}\label{eqn:SaddlePoints_largeeta2}
    (\rho_0, \theta)_{*} = 
        \left(1, \pm\mathrm{atan}\left[ \frac{\sqrt{\eta^2 - 4(1+\tilde{q})^2}}{2(1+\tilde{q})} \right] + 2n\pi \right) .
\end{equation}
For values of $\tilde{q}$ beyond these regimes, the separatrix vanishes and the phase portrait is entirely consumed by the ZD phase. 



\subsection{Dynamical phase transitions}
The separation of the phase portrait into ZD and ID phases can be probed by studying the quench dynamics of a fixed initial state as a function of the normalized Zeeman shift $\tilde{q}$ and fixed $\eta$. In Fig.~\ref{fig:DynamicalPhaseDiagram}(a) we show some example dynamics for the initial condition $\rho_0(0) = 1/2$ and $\theta(0) + \varphi_j = 0$ with relative interaction strength $\eta = 0.2$ [row i)] and $1.8$ [row ii)]. For relatively small values of $\tilde{q}$ (solid red lines) the dynamics is in the ID regime, as indicated by the fact that $\cos(\theta/2)\geq 0$ throughout the dynamics. On the other hand, when $\tilde{q}$ is relatively large (dashed blue lines) we find $\cos(\theta/2)$ oscillates between the extreme values of $\cos(\theta/2) = \pm 1$. Motivated by this we define a time-averaged order parameter, 
\begin{equation}
    O_{\theta} = \overline{\cos(\theta/2)} = \lim_{T\to\infty} \frac{1}{T} \int_0^T ~\cos(\theta(t)/2) ~dt ,
\end{equation}
which distinguishes the dynamical phases and features non-analytic behaviour at a critical value of $\tilde{q} = \tilde{q}_c$ separating them, i.e., a dynamical phase transition. Specifically, the ID regime corresponds to an ordered phase with $O_{\theta} \neq 0$, while the ZD regime is disordered and distinguished by $O_{\theta} = 0$. 

We note that the dynamics of the fractional population appear qualitatively similar in both phases, in the sense that $\rho_0$ exhibits periodic oscillations for both large and small values of $\tilde{q}$. Thus, despite the fact that $\rho_0$ is more straightforward to access experimentally than the spinor phase $\theta$ \cite{Yingmei2009qpt}, a quantity such as the time-averaged value $\overline{\rho}_0$ is in general inappropriate to use as an order parameter. Nevertheless, finer features of these oscillations, such as the period, do display fingerprints that can be used to diagnose the phases and the DPT separating them \cite{Austin2024,Zhang2005,Duan_DPT_2019}. For example, at the boundary between the ID and ZD phases [see dotted red lines in Fig.~\ref{fig:DynamicalPhaseDiagram}(a)], the dynamics drastically slows down and the oscillation amplitude of $\rho(t)$ is maximal. More generally, within our model the behavior of the amplitude is also closely related to the distinct locations of the fixed points through which the separatrix passes as a function of $\eta$. Specifically, for $\eta = 1.8$ the fractional population asymptotically approaches $\rho = 0$ when $\tilde{q}$ is tuned close to the DPT due to the location of the fixed point described by Eq.~(\ref{eqn:SaddlePoints_largeeta1}), while for $\eta = 0.2$ the population approaches a fixed point with non-zero $\rho_0 = (2-2\tilde{q}-\eta)/(4-2\eta) \approx 0.23 $ [see Eq.~(\ref{eqn:SaddlePoints_smalleta})].

The location of the DPT in parameter space can be obtained by equating the mean-field energy $E_0$ of the fixed initial condition with that of the separatrix trajectory defined by Eqs.~(\ref{eqn:SaddlePoints_smalleta})-(\ref{eqn:SaddlePoints_largeeta2}) that delineates the ID and ZD regimes. Another approach, though fundamentally equivalent, is to 
develop the dynamical phases and transition from a reduced one-dimensional ($1$D) model of the dynamics, analogous to the description of the motion a fictitious classical particle in a potential well \cite{Jamir2022,RLS_Dicke_2021}. Specifically, by using conservation of energy we can reduce the pair of mean-field equations of motion [Eq.~(\ref{eqn:classicalEOM})] to a single differential equation of the form $(\hbar\dot{\theta}/\mathcal{G}_0)^2 + \mathcal{V}(\theta)/\mathcal{G}_0 = 0$. 
Here, $\dot{\theta}$ plays the role of the velocity of the fictitious particle, which is defined by position coordinate $\theta$ and moves in the effective potential well described by $\mathcal{V}(\theta)$. Expressions for $\mathcal{V}(\theta)$ are given in Appendix~\ref{app:potentials}.

The effective potential depends on the choice of fixed initial condition, but always features a periodic multi-well structure as shown in a few characteristic examples in Fig.~\ref{fig:DynamicalPhaseDiagram}(b). There we fix $\rho_0(0) = 1/2$ and $\theta(0) + \varphi_j = 0$ and consider a few relevant values of $\tilde{q}$ and $\eta$ corresponding to the different dynamical phases. In the language of the effective $1$D potential, these phases arise due to the fact that the total mechanical energy of the fictitious particle is zero and thus the motion within the potential must be constrained to regions where $\mathcal{V}(\theta)\leq0$. As a result, in the ID phase [see panels i) and iv)], interactions lock the spinor phase to remain between $-\pi < \theta(t) < \pi$, which is manifested through the confinement of the fictitious particle to a single well of the potential surrounded by inaccessible regions where $\mathcal{V}(\theta) > 0$. On the other hand, in the ZD phase [see panels ii) and iii)] the potential is negative everywhere and thus the particle can explore all values of $\theta$.

At a fixed value of $\eta$, the location of the DPT in terms of $\tilde{q}_c$ is obtained by finding when the roots of $\mathcal{V}(\theta)$ correspond to the maxima of the effective potential, i.e., when the dynamics transitions between bounded and unbounded growth of the spinor phase. 
For simplicity, we restrict our detailed treatment of the DPT to a pair of cases where the initial spinor phase and phase offset are chosen to together satisfy: i) $\theta(0) + \varphi_{j} = 0$ or ii) $\theta(0) + \varphi_{j} = \pi$. 

For the first scenario of $\theta(0) + \varphi_{j} = 0$, the critical Zeeman energies are themselves broken into subcases depending on the relative interaction strength. 
Defining $\eta_{c,+} = 2\rho_0/(2-\rho_0)$ and $\eta_{c,-} = 2(1-\rho_0)/(1+\rho_0)$, we find: i) for $\eta < \eta_{c,\pm}$, 
\begin{equation}\label{eqn:qc_case1}
    \tilde{q}_{c,\pm} = \frac{1}{2}[1-2\rho_0(0)](2 - \eta) \pm \sqrt{2\eta(2-\eta)\rho_0(0)[1-\rho_0(0)]} ,
\end{equation}
and ii) for $\eta \geq \eta_{c,\pm}$ [the separatrix passes through the fixed points defined by Eqs.~(\ref{eqn:SaddlePoints_largeeta1}) or (\ref{eqn:SaddlePoints_largeeta2})],
\begin{equation}\label{eqn:qc_case2}
    \tilde{q}_{c,\pm} = \frac{1}{4}[(2+\eta)(1-2\rho_0(0) \pm 1)] .
\end{equation}
The ID dynamical phase exists for $\tilde{q}_{c,-} \leq \tilde{q} \leq \tilde{q}_{c,+}$ and the system exhibits ZD dynamics otherwise. Figure \ref{fig:DynamicalPhaseDiagram}(b) illustrates the structure of the predicted dynamical phase diagram as a function of $\eta$ and $\tilde{q}$ for the initial condition $\rho_0(0) = 1/2$ and $\theta(0)+\varphi_j = 0$.

For the second scenario of $\theta(0) + \varphi_{j} = \pi$ there is a unique value of  
\begin{equation}\label{eqn:qc_case3}
    \tilde{q}_c = \frac{1}{2}(2+\eta)[1-2\rho_0(0)] ,
\end{equation}
regardless of $\eta$. This relative Zeeman shift marks a singular point where the system is prepared exactly at an unstable fixed point in the classical phase space, and separates neighboring regimes of ZD dynamics. 


The delineation of the critical Zeeman shift into a number of subcases dependent on $\eta$ is directly related to the nature of the separatrices in the phase portrait of Fig.~\ref{fig:PhasePortrait} [see also Eqs.~(\ref{eqn:SaddlePoints_smalleta})-(\ref{eqn:SaddlePoints_largeeta2})], or equivalently the structure of the effective potential $\mathcal{V}(\theta)$. Inspection of the callouts in Fig.~\ref{fig:DynamicalPhaseDiagram}(b) demonstrates that for $\eta > \eta_{c,+}$ [panels i) and ii)] an additional pair of wells emerge in the potential [relative to the case of $\eta < \eta_{c,+}$ in panels iii) and iv)]. While this does not change the qualitative description of the system in terms of ID and ZD phases, it does affect the details of the dynamics close to the proximity of the critical Zeeman energy (i.e., $\tilde{q} \approx \tilde{q}_{c,+}$). We expect that this can be a particularly interesting direction to explore in the context of the quantum spin-mixing dynamics \cite{SpinorQFI}, as well as the growth of quantum fluctuations and entanglement near DPTs \cite{RLS_Dicke_2021,Qingze2021,SpinorQFI}.



\begin{figure}[tb]
    \centering
    \includegraphics[width=8cm]{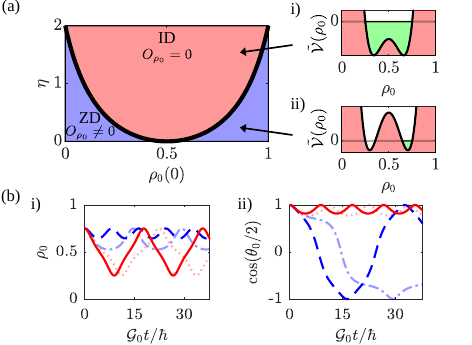}
    \caption{(a) Dynamical phase diagram with fixed $\tilde{q} = 0$, characterized by $O_{\rho_0} = \overline{\rho}_0-1/2$ (see text), as a function of relative interaction strength $\eta$ and initial condition $\rho_0(0)$ with fixed $\theta(0) + \varphi_j = 0$. Call-outs i) and ii) show the effective $1$D potential $\tilde{\mathcal{V}}(\rho_0)$ (plotted in units of $\mathcal{G}_0$) for indicated parameter regimes (see Appendix~\ref{app:potentials}). The green and red shaded regions indicate the accessible and inaccessible values of $\rho_0$ during dynamics, respectively. Faded horizontal line in each plot indicates $\tilde{\mathcal{V}}(\rho_0) = 0$ to highlight the turning points of the dynamics. 
    (b) Example time-traces for the conjugate variables $(\rho_0,\theta)$ for i) $\eta = 0.2$ and ii) $\eta = 1.8$ (same as phase portraits in Fig.~\ref{fig:PhasePortrait}). We compare dynamics for $\eta/\eta_c = 0.7$ (red solid lines), $0.99$ (faded red dotted lines), $1.05$ (faded blue dot-dashed lines) and $1.15$ (dashed blue lines). 
    } 
    \label{fig:DynamicalPhaseDiagram_q0}
\end{figure}

Complementary to the above analysis, and unique to the driven model featuring tunable $\eta$, one can realize a DPT even when the effective Zeeman energy is fixed to $\tilde{q}=0$. The dynamical phase diagram is illustrated in Fig.~\ref{fig:DynamicalPhaseDiagram_q0}(a) as a function of $\eta$ and initial fractional population $\rho_0(0)$ [we fix $\theta(0) + \varphi_{j} = 0$ in the following]. A critical relative interaction strength,
\begin{equation}
    \eta_c = \frac{2(1-2\rho_0)^2}{1+4\rho_0 - 4\rho_0^2} , 
\end{equation}
delineates an ID phase ($\eta > \eta_c$) from a ZD phase ($\eta < \eta_c$). In contrast to the prior examples with $\tilde{q}\neq 0$, this dynamical phase diagram can be characterized through the time-averaged behavior of the fractional population $\rho_0$ with order parameter 
\begin{equation}
    O_{\rho_0} = \overline{\rho}_0 - \frac{1}{2} = \lim_{T\to\infty} \frac{1}{T} \int_0^T ~\left[\rho_0(t) - \frac{1}{2}\right] dt .
\end{equation}
To gain insight into this distinction to $\tilde{q} \neq 0$, in the call-outs of Fig. \ref{fig:DynamicalPhaseDiagram_q0}(a) we illustrate the quartic effective potential $\tilde{\mathcal{V}}(\rho_0)$ associated with $\rho_0$ (see Appendix~\ref{app:potentials} for details), which distinguishes a pair of dynamical phases: i) an ordered phase for $\eta < \eta_c$ with $O_{\rho_0} \neq 0$, in which $\rho_0$ is confined to a single well of the potential (ZD phase) and ii) a disordered phase for $\eta > \eta_c$ with $O_{\rho_0} = 0$ where both wells are traversed (ID phase). We emphasize that the ordered and disordered phases in terms of $O_{\rho_0}$ are reversed with respect to the ID and ZD phases defined by $O_{\theta}$ [compare the timetraces in Fig.~\ref{fig:DynamicalPhaseDiagram_q0}(b)].

The quartic effective potential and the use of the fractional population to define the order parameter $O_{\rho_0}$ can be identified as analogous to studies of macroscopic self-trapping DPTs in BECs \cite{Smerzi1997selftrapping,Oberthaler2005selftrapping} and the associated Lipkin-Meshkov-Glick model \cite{Muniz2020} upon identification of the ordered phase with the ``trapped'' regime (and disordered with ``untrapped''). These phases can be similarly characterized using the dynamics of the population imbalance or magnetization, respectively. We note that this connection is unique to our model with tunable $\eta$: The ``standard'' DPT of a spinor BEC \cite{Zhang2005}, which can be recovered by setting $\eta = 2$, is instead associated with a cubic effective potential for $\rho_0$ \cite{Zhang2005} that does not support a trapped-untrapped transition characterized through a time-averaged function of the fractional population.

\section{Quench dynamics beyond the effective model\label{sec:GPE}}
In the previous sections we have developed a spin-only description for the spin-mixing dynamics in the presence of a periodically modulated external potential. Here, we present a theoretical investigation of an example implementation of such a system based on a spin-$1$ BEC in a periodically driven harmonic potential. In particular, we carry out detailed Gross-Pitaevskii (GP) simulations that captures both spatial and spin dynamics to assess the validity of the dSMA, our simplified few-mode model, and the robustness of the predicted spin-mixing dynamical phase diagram.
We anticipate that a broader range of $\eta$ can be achieved by applying methods similar to what is considered below to different systems, such as traps with different geometries and lattice confinement.
To access the entire dynamical phase diagram, 
one would look at other implementations where 
$\eta$ can vary over a wide range. 
For the exemplar system discussed below, we have access to $\eta\le 0.23$.


\subsection{Example setup and Gross-Pitaevskii model}
For our exemplar system, we consider a harmonic confining potential that is periodically driven according to 
Eq.~\eqref{trap_potential} and Eq.~\eqref{trap_freq}.
To explore the tunability of the dynamical phase diagram via the driving phase $\varphi$ of the modulated potential, 
we will consider the cases $\varphi = 0$ and $\varphi = \pi$, which we will refer to as Type I and Type II driving, respectively. To be concrete, we choose the total condensate particle number to be $N = 10^3$ and a background trap frequency of $\omega_{\text{osc},0} = 2\pi \times 1.225$~kHz, which can be realized using a chip trap~\cite{Hansel2001,Hansch1999}. We base our simulations on a condensate of $^{23}$Na atoms, such that $g_0 = 0.638 \times 10^{-1} E_{\text{ho}} a_{\text{ho}}^3$ and $g_2 = 0.227 \times 10^{-2} E_{\text{ho}} a_{\text{ho}}^3$, where the energy and length units are  
$E_{\text{ho}} = \hbar\omega_{\text{osc},0}$ and $a_{\text{ho}} = \sqrt{\hbar/(M\omega_{\text{osc},0})}$, respectively.

The driving frequency $\omega$ and amplitude $\omega_{\text{osc},1}$ require careful selection, consistent with the discussion in Sec.~\ref{sec:model}. On the one hand, $\omega$ should be sufficiently large so that the spin-mixing resonances described by Eq.~(\ref{eqn:H3mode}) are far from the energy scales of the background spin-mixing Hamiltonian determined by $\mathcal{G}_0$. On the other hand, the value of $\omega$ should be kept sufficiently small to ensure that the spatial degrees of freedom are driven effectively adiabatically. This latter adiabaticity requirement implies that the tunability of $\omega$ and $\omega_{\text{osc},1}$ depend on each other, as a larger driving amplitude increases the likelihood of creating undesirable spatial excitations. In this work, we choose $\omega=2\pi\times 0.6$~kHz and limit the value of $\omega_{\text{osc},1}$ to at most $2\pi\times 0.707$~kHz. 

We model the coupled dynamics of the spin and spatial degrees of freedom by numerically integrating the time-dependent spinor GP equation (GPE) \cite{kawaguchi_2012,StamperKurn2013spinor},
\begin{widetext}
\begin{align}
\label{eqn:spinorGP} i\hbar\frac{\partial}{\partial t}\begin{pmatrix}
\psi_{-1}\\
\psi_0\\
\psi_1
\end{pmatrix}& =
\left[-\frac{\hbar^2\nabla^2}{2M} + V(\mathbf{r}, t) +
g_0(N-1)\left(|\psi_{-1}|^2+|\psi_{0}|^2+|\psi_{1}|^2\right)\right]
\begin{pmatrix}
\psi_{-1}\\
\psi_0\\
\psi_1
\end{pmatrix}
+\nonumber
\begin{pmatrix}
q & 0 & 0\\
0 & 0 & 0\\
0 & 0 & q
\end{pmatrix}
\begin{pmatrix}
\psi_{-1}\\
\psi_0\\
\psi_1
\end{pmatrix}\\\nonumber
& + g_2(N-1)\begin{pmatrix}
|\psi_{-1}|^2+|\psi_{0}|^2-|\psi_{1}|^2 & \psi_1^*\psi_0 & 0\\
\psi_1\psi_0^* & |\psi_{1}|^2+|\psi_{-1}|^2 & \psi_{-1}\psi_0^*\\
0 & \psi_{-1}^*\psi_0 & |\psi_{1}|^2+|\psi_{0}|^2-|\psi_{-1}|^2
\end{pmatrix}
\begin{pmatrix}
\psi_{-1}\\
\psi_0\\
\psi_1
\end{pmatrix}.
\\
\end{align}
\end{widetext}
This GPE corresponds to a mean-field treatment of the dynamics described by Eqs.~(\ref{eqn:SpatialHam}) and (\ref{eqn:SpinHam}), i.e., it includes contributions from both the spin-independent ($g_0$) and spin-dependent $(g_2)$ interactions, as well as a confining potential $V(\mathbf{r}, t)$. Within the GP framework, the Zeeman components of the spinor BEC are fully described by the GP wave functions $\psi_m(\mathbf{r}, t)$, which satisfy the normalization condition $\int \left(|\psi_{-1}|^2 + |\psi_0|^2 + |\psi_1|^2\right)d^3\mathbf{r}=1$.

Results in this section are obtained by numerically integrating Eq.~(\ref{eqn:spinorGP}) using the software package \textit{XMDS2} and the in-built Runge-Kutta propagator \textit{RK4}~\cite{Dennis2013201}. 
We enforce that each spin component remains spherically symmetric during the entire dynamics. 
While enforcing spherical symmetry throughout the dynamics may miss some dynamics associated with the angular degrees of freedom, 
we expect that our treatment captures the essential dynamics.
We compute the kinetic energy contribution to Eq.~(\ref{eqn:spinorGP}) via the zeroth spherical-Bessel transformation.
For all parameter regimes considered in this work, 
we have checked that a radial box size of $r_{\text{max}}=7.35a_{\text{ho}}$
with 128 spatial grid points 
and a time step of  
$\Delta t=0.471\times10^{-3}\hbar/E_{\text{ho}}$  
ensures the convergence of all numerical results.

\begin{figure}
    \includegraphics[width=0.4\textwidth]{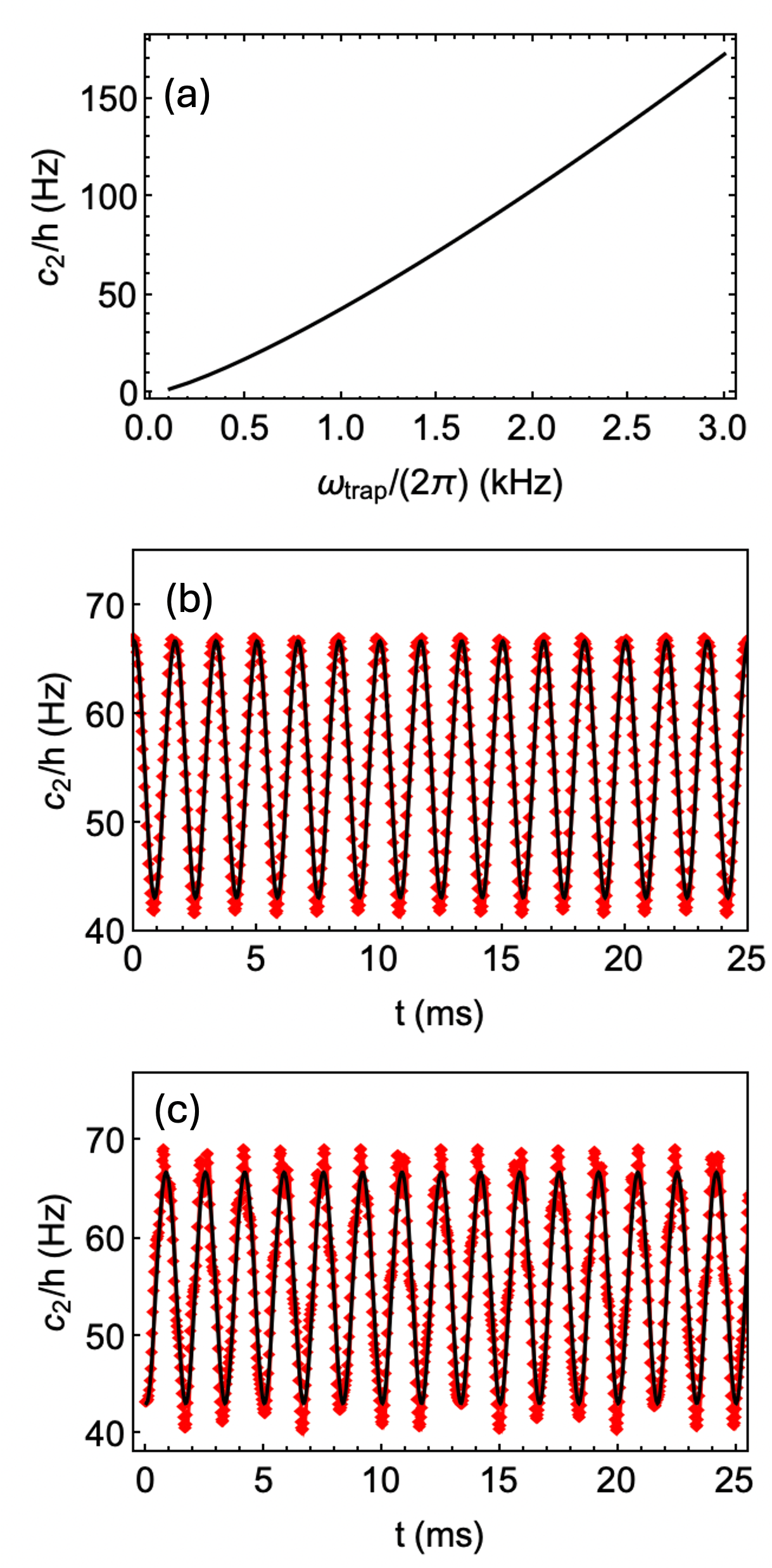}
    \caption{(a) The value of $c_2$ for the $^{23}$Na BEC in a 3D harmonic trap with frequency $\omega_{\text{trap}}$ and particle number $N=10^3$.
    (b) and (c) The value of $c_2$ for Type I and Type II periodically driven trapping potentials (see text). The black line and red diamonds correspond to the adiabatic limit and the numerical GP simulations, respectively.
    The results are for a scalar BEC with $100\%$ population in the $m=+1$ state.}
    \label{fig:c2}
\end{figure}

To assess the adiabaticity of the driving protocol, we calculate the ground state wavefunction using imaginary time propagation of the GPE (\ref{eqn:spinorGP}) for a scalar BEC with $100$\% population in the $m=+1$ state. We use the ground state wavefunction to compute the associated spin-dependent interaction strength $c_2$ according to Eq.~(\ref{eqn:c2_t}) at various static trap frequencies $\omega_{\text{trap}}$. The resulting values of $c_2$ are shown in Fig.~\ref{fig:c2}(a).
We then compare the time-dependent values of $c_2(t)$ obtained by numerical simulation of the driven potential according to the time-dependent GPE for an identical scalar $m=+1$ BEC (red symbols) to the ground state results for $c_2$ obtained at the same instantaneous trap frequency $\omega_{\text{trap}}(t)$ (black lines) in Figs.~\ref{fig:c2}(b) and~\ref{fig:c2}(c) for Type I and Type II driving, respectively. In both panels we use $\omega_{\text{osc},1}=2\pi\times 0.707$~kHz to modulate the harmonic potential. We observe that the behaviour of the spin-dependent interaction is ``nearly" adiabatic, in the sense that only small deviations are observed near the maximum/minimum values of $c_2$ while the periodicity of the interaction is clearly inherited from the underlying modulation. We note that the Type II driving scheme appears to show slightly larger quantitative disagreement than Type I driving. For the parameters considered here and the adiabatic driving of the two types, $c_2$ oscillates between $h\times 43.12$~Hz and $h\times 66.78$~Hz while the chemical potential (not shown here) oscillates between $h\times 3.00$~kHz and $h\times 4.48$~kHz, i.e., the chemical potential is much larger than $c_2$, $\omega_{\text{osc}, 1}$, and $\omega$. 



\begin{figure*}[ht!]
    \centering
    \includegraphics[width=1\textwidth]{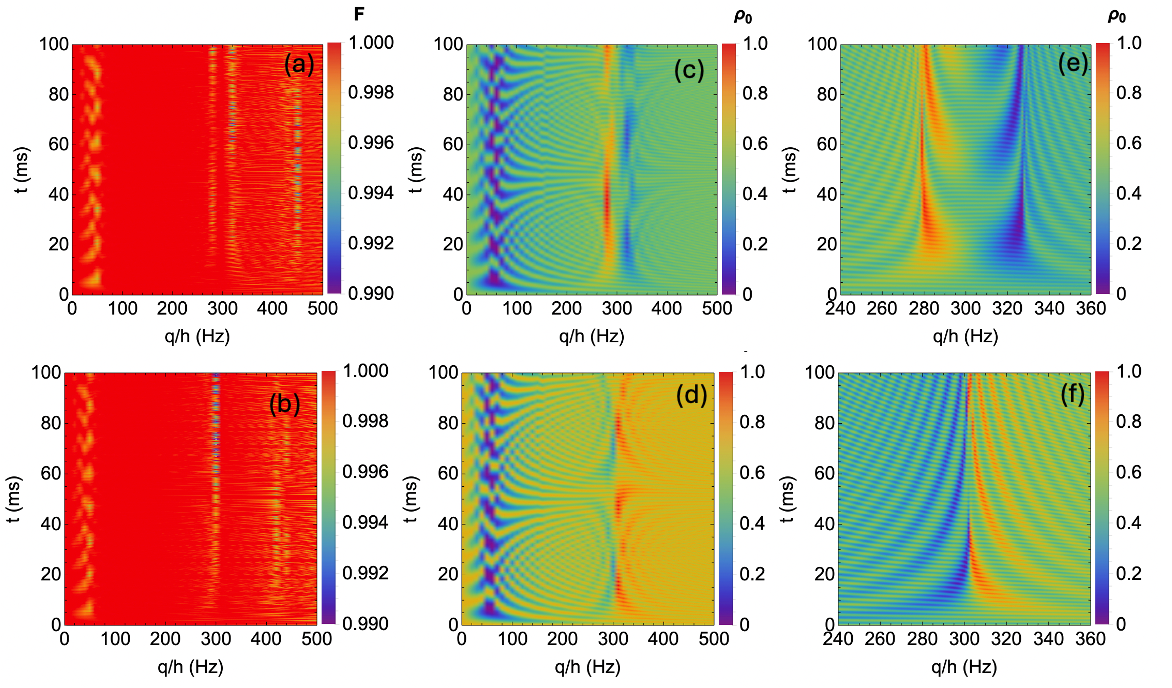}
    \caption{Time evolution of state fidelity $F$ and fractional population of $\rho_0$ over a range of Zeeman shifts $q$ encompassing expected resonances. The first and second rows correspond to Type I and Type II driving results, respectively. Panels (e) and (f) zoom in on Panels (c) and (d) around the resonance positions. The driving frequency is $\omega = 2\pi \times 0.6$~kHz, with driving amplitude $\omega_{\text{osc},1} = 2\pi \times 0.707$~kHz. The initial state is characterized by $\rho_0 = 0.5, \langle\hat{M}\rangle=0$, and $\theta=0$.
}
    \label{fig:spin_fidelity_map}
\end{figure*}

Before attempting to quantitatively compare the results of 
the GP model
to the effective time-dependent spin model we developed in Sec.~\ref{sec:model}, 
we demonstrate the validity of the underpinning dSMA~\cite{Hardesty2023}, i.e., that the condensate occupies a single spatial mode during the dynamics. In particular, the dSMA may break down as the system evolves for a number of reasons, including accidental resonances between the Zeeman energy $q$ and spatial excitations of the condensate~\cite{Jie2020, Jianwen2023}.
To quantitatively assess the validity of the dSMA, 
we thus compute the time-dependent, normalized state overlap between the $m=-1$ and the $m=0$ components:
\begin{eqnarray}
F=\frac{|\langle\psi_{-1}|\psi_0\rangle|}{\sqrt{\langle\psi_{-1}|\psi_{-1}\rangle\langle\psi_0|\psi_0\rangle}},
\end{eqnarray}
where $\langle \psi_m \vert \psi_{m'} \rangle = \int \psi^*_m \psi_{m'} ~d^3\mathbf{r}$. Note that the overlap between the $m=1$ and the $m=0$ components, which can be calculated in a similar way, leads to the same result since we always work with the initial states with $\langle\hat{M}\rangle=0$, for which the densities of the $m=\pm 1$ components remain identical during the dynamics. 
Figures~\ref{fig:spin_fidelity_map}(a) and~\ref{fig:spin_fidelity_map}(b) show the time trace of $F$ using an initial state with $\rho_{0}=0.5$, $\langle\hat{M}\rangle=0$, $\theta=0$ and $\omega_{\text{osc},1}=2\pi\times 0.707$~kHz for a range of $q$. 
The value of $F$ maintains above $0.99$ throughout for these two types of drivings, which confirms the validity of dSMA for all the $q$ values considered here. 

\subsection{Dynamical phase diagram}
Figures~\ref{fig:spin_fidelity_map}(c) and~\ref{fig:spin_fidelity_map}(d) show the time traces of the fractional population $\rho_0$ as a function of the Zeeman energy $q$ for Type I and II driving, respectively. Here, and throughout the results presented in the remainder of this section, we fix $\theta(0) = 0$ by choosing the initial GP wavefunctions for each Zeeman component to have identical phases. 
Two clear features are observable in both panels (c) and (d). First, a ``valley'' around $q/h \approx 50$~Hz, which is characterized by slow oscillations of $\rho_0$ that approach zero. This feature is a signature of spin-mixing dynamics described by the ``background'' contribution ($\propto \mathcal{G}_0$) of the driven few-mode model Eq.~(\ref{eqn:DrivenH3mode}) \cite{Zhang2005}. 
The second feature are the additional resonances located near $q/h \approx 300$~Hz, i.e., at about half the driving frequency $\omega$; these resonances are consistent with the predicted effects of the modulated confining potential according to our effective model Eq.~(\ref{eqn:Heff}) and Sec.~\ref{sec:DynamicalPhaseDiagram}.
For Type I driving [panel (b)], we observe
two resonances signaled by slow spin-mixing dynamics at $q/h \approx 280$~Hz and $q/h \approx 330$~Hz [see panel (e) for a zoom-in of this region]. 
For Type II driving [panel (d)], 
a single resonance appears at $q/h = 300$~Hz [see panel (f) for a zoom-in of this region]. 

\begin{figure*}[ht!]
    \centering
    \includegraphics[width=1.0\textwidth]{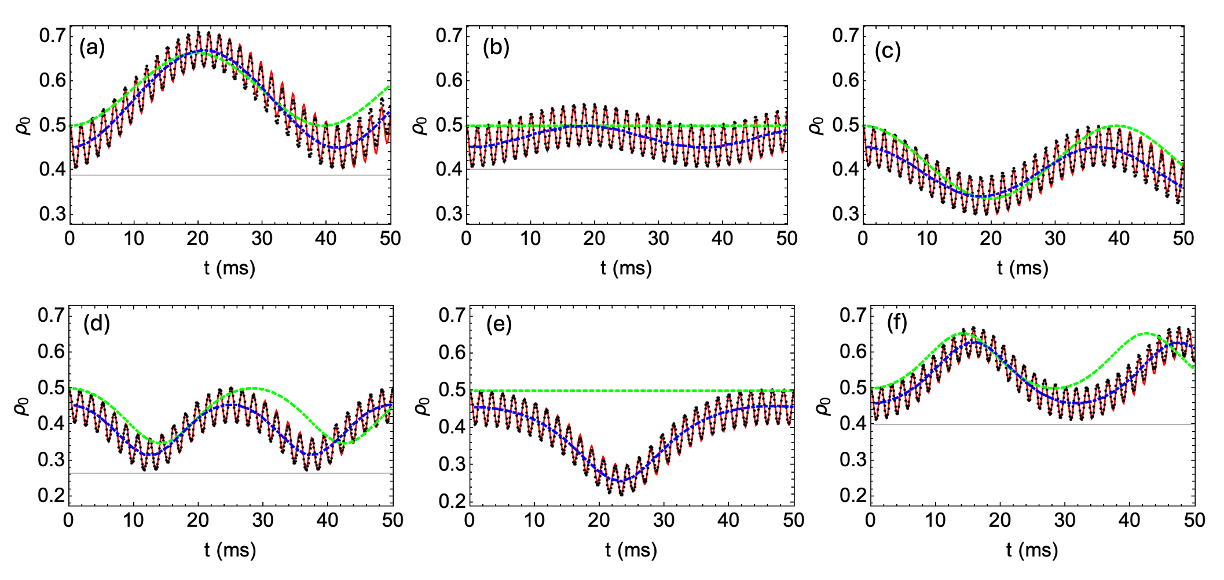}
    \caption{Comparison of results from the GP simulation (black dots), the effective model $H_{\text{mf},1}$ (green dashed line), and dSMA simulation (red solid line) for (a,d) $q/h = 290$~Hz (below resonance), (b,e) $q/h = 300$~Hz (near resonance), and (c,f) $q/h = 310$~Hz (above resonance). The first and second rows correspond to Type I and Type II driving cases, respectively. The blue dot-dashed lines show the GP results after smoothing the fast oscillations (see main text).
    }
    \label{fig:GP_model_comparison}
\end{figure*}


Next, we obtain the underlying time-dependent interaction $c_2(t)$ for our effective few-state model. 
Expanding $c_2(t)$ as a cosine series according to Eq.~\eqref{eqn:c2_ansatz}, we compute the expansion coefficients via, 
\begin{equation}
\mathcal{G}_k =\frac{\int_0^{t_{\text{max}}} c_2(t) \cos(k\omega t-\varphi_k)dt}{\int_0^{t_{\text{max}}} \cos^2(k\omega t-\varphi_k)dt}.
\end{equation}
We have $\varphi_0=0$. 
The Type I and Type II driving sequences we have defined above correspond to the cases $ \varphi_1 = 0$ and $\varphi_1 = \pi$, respectively. 
For an initial phase $\theta(0)=0$, we therefore have
$\theta(0) + \varphi_1=0$ and $\theta(0) + \varphi_1=\pi$,
which correspond to the cases discussed in Sec.~\ref{sec:DynamicalPhaseDiagram}.
Using $t_{\text{max}} = 100$~ms (which is large enough to achieve converged results), 
we obtain $\mathcal{G}_0 = h \times 55.31$~Hz and $\mathcal{G}_1 = h \times 11.80$~Hz for $\omega_{\text{osc},1} = 2\pi \times 0.707$~kHz for Type I and Type II driving sequences.
We have checked that the magnitudes of $\mathcal{G}_k/h$ with $k>1$ are less than $0.1$~Hz; they thus can be safely neglected.

\begin{figure*}[ht!]
    \centering
    \includegraphics[width=1.0\textwidth]{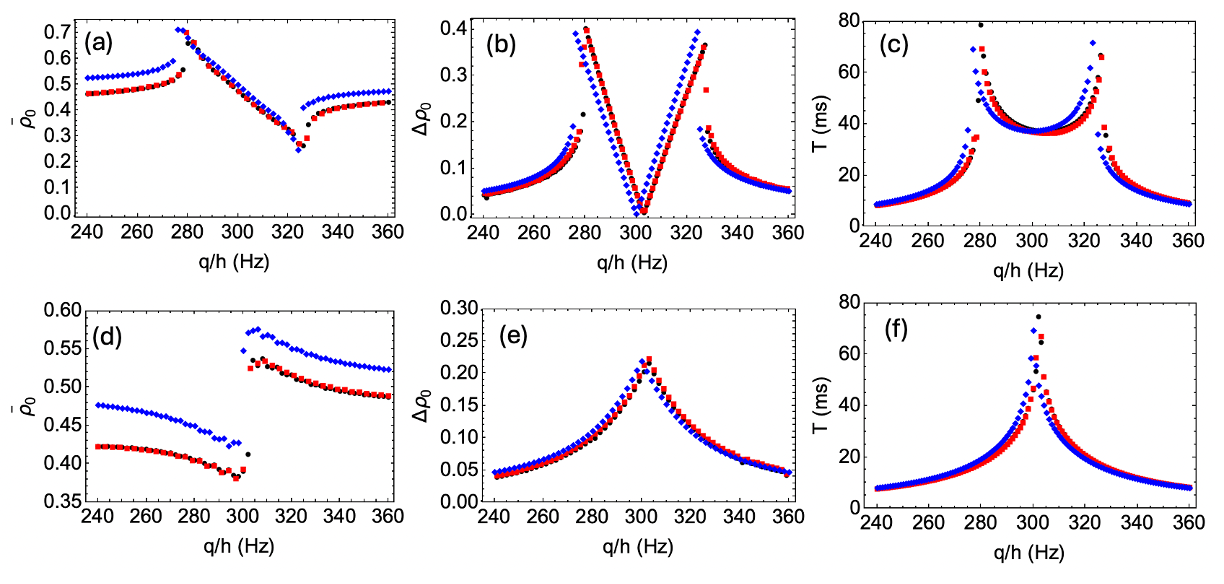}
    \caption{Time-averaged fractional population $\bar{\rho}_0$, oscillation visibility $\Delta \rho_0 = \rho_{0,\text{max}} - \rho_{0,\text{min}}$, and spin oscillation period $T$ for Type I driving (first row) and Type II driving (second row). Black circles, blue diamonds, and red squares correspond to results from the smoothed GP model, effective model $H_{\text{mf}, 1}$, and the dSMA model, respectively.}
    \label{fig:period_order_parameter}
\end{figure*}

Figure~\ref{fig:GP_model_comparison} shows a comparison of specific timetraces of $\rho_0$ obtained from the GP calculations (black dots), the dSMA model (\ref{eqn:DrivenH3mode}) using $c_2(t) = \mathcal{G}_0 + \mathcal{G}_1 \cos(\omega t -\varphi_1)$ and the time-independent few-mode model $H_{\text{mf},1}$ [Eq.~(\ref{eqn:Hmf})], which only retains the near-resonant contribution. We show results for Zeeman energies close to $q = \hbar\omega/2$ for Type I (top panels) and Type II (bottom panels) driving. 
In all cases, the dSMA results agree near perfectly with the full GP simulations. 
Since the derivation of the effective model $H_{\text{mf},1}$ is equivalent to integrating out fast oscillations, 
we further smooth the GP results by averaging over each oscillation cycle, i.e.,
all points within two adjacent peaks of the fast oscillation
(black solid lines in Fig.~\ref{fig:GP_model_comparison}). 
The smoothed curves are useful for extracting the period and amplitude of the slow spin oscillations (see later results shown in Fig.~\ref{fig:period_order_parameter}).  
Note that the method used to average out these fast oscillations is not unique; 
alternatives include connecting adjacent valleys or peaks, which can cause minor shifts in the relative location of the smoothed $\rho_0$ curve. 
However, the characteristic shape of the smoothed curve and the location of the extracted resonance as a function of $q$ remain unaffected. 
We observe that the predictions of the effective few-mode model aligns well with smoothed GP results for $q$ below and above resonance [Figs.~\ref{fig:GP_model_comparison}(a),~\ref{fig:GP_model_comparison}(c),~\ref{fig:GP_model_comparison}(d), and~\ref{fig:GP_model_comparison}(f)], 
with slight deviations attributable to the off-resonant (i.e., background) spin-changing collision terms proportional to $ \mathcal{G}_0$ that are neglected. 
This discrepancy is most evident exactly at $q/h =  0.3$~kHz, where the effective model predicts a static $\rho_0 = 1/2$, 
while the results of the smoothed GP data reveal small residual oscillations [Figs.~\ref{fig:GP_model_comparison}(b) and~\ref{fig:GP_model_comparison}(e)].

To further benchmark the effective description of the spin-mixing dynamics, we map out the dynamical phase transition as a function of the Zeeman energy through signatures accessible from the fractional population. Specifically, we analyze the time-averaged fractional population $\bar{\rho}_0$, the oscillation visibility $\Delta \rho_0 = \rho_{0,\text{max}} - \rho_{0,\text{min}}$ (where $\rho_{0,\text{max/min}}$ are the maximum and minimum values of $\rho_0$ over time), and the oscillation period $T$ for Zeeman energies near the resonance $q \approx \hbar\omega/2$. 
The results for these metrics are shown in Fig.~\ref{fig:period_order_parameter} for both Type I (top row) and Type II (bottom row) driving according to the GPE simulations (black circles), dSMA model (blue diamonds) and time-independent effective model $H_{\text{mf},1}$ (red squares). In all cases, we obtain the metrics from simulations integrating up to $t_{\text{max}} = 100$~ms.  
For the GPE and dSMA model, fast oscillations in $\rho_0(t)$ are smoothed out before computation of all metrics. 
As discussed above, depending on the smoothing method, 
the $\bar{\rho}_0$ curves for the GP and dSMA models are vertically shifted relative to the effective model. 
However, key features such as cusp structures in the Type I driving data and discontinuities near the resonance in the Type II driving data agree well with the effective model [see Figs.~\ref{fig:period_order_parameter}(a) and~\ref{fig:period_order_parameter}(d)]. 
Similarly, the oscillation visibility $\Delta \rho_0$ and period $T$ show consistent agreement between all three models, up to a slight shift of around $2$~Hz in the positions of the resonances observed in the GPE and dSMA data due to the off-resonant spin-changing collisions (i.e., terms driven by $\mathcal{G}_0$).


\begin{figure}
    \centering
    \includegraphics[width=0.45\textwidth]{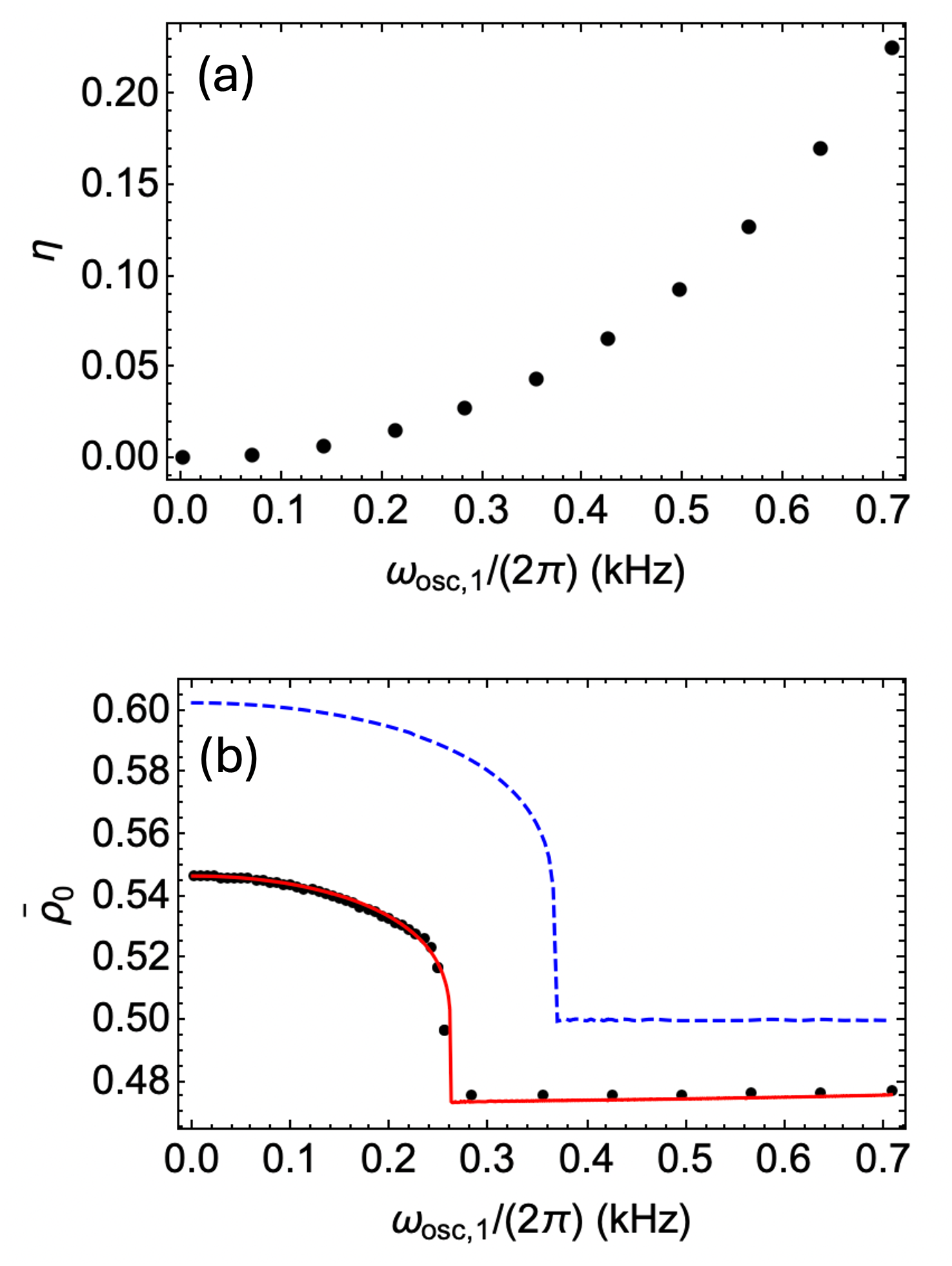}
    \caption{Dynamical phase transition for $q = \hbar \omega/2$ and $\varphi_1 + \theta = 0$. (a) $\eta$ as a function of the driving amplitude $\omega_{\text{osc},1}$. (b) Time-averaged population fraction $\bar{\rho}_0$ for the initial state with $\rho_0= 0.6, \langle\hat{M}\rangle=0, \theta=0$ and varying $\omega_{\text{osc},1}$. The black dots, red solid line, and blue dashed line in (b) correspond to the GP result, the dSMA result, and the effective model $H_{\text{mf},1}$ result, respectively. All results use $\omega = 2\pi \times 0.6$~kHz and $\omega_{\text{osc},0} = 2\pi \times 1.225$~kHz.
}
    \label{fig:dpt_rho0}
\end{figure}

To observe the dynamical phase transition at $q_{\mathrm{eff},1} = 0$ ($q = \hbar\omega/2$), which is supported by the time-independent effective model (see Fig.~\ref{fig:DynamicalPhaseDiagram_q0} and surrounding discussion), we focus on Type I driving and perform GP simulations for an initial state with $\rho_0 = 0.6, \langle\hat{\mathcal{M}}\rangle = 0$, and $\theta = 0$. By scanning the driving amplitude $\omega_{\text{osc},1}$ of the harmonic potential we are able to tune the interaction ratio $\eta$.
Figure~\ref{fig:dpt_rho0}(a) shows $\eta$ as a function of $\omega_{\text{osc},1}$, indicating that we achieve a maximum value of $\eta \approx 0.23$, which is more than sufficient for this investigation.
In addition, we find that $\mathcal{G}_{k>1}$ remains less than $1\%$ of $\mathcal{G}_1$ for all values of $\omega_{\text{osc},1}$ considered here.
Figure~\ref{fig:dpt_rho0}(b) shows a comparison of the extracted $\bar{\rho}_0$ obtained from the time-independent effective model (orange dashed line), GPE simulations (blue dots), and the dSMA model (black solid line). We observe that the overall structure of the transition of the effective model is preserved in the GPE and dSMA results, but the critical point is shifted to a lower driving amplitude $\omega_{\text{osc},1}$, i.e., smaller $\eta$, and the observed $\bar{\rho}_0$ is shifted to smaller values. We attribute this to a combination of effects from the smoothening of the underlying GPE and dSMA data as well as the influence of the off-resonant spin-changing collisions that the simpler effective model neglects. In principle, all of these effects can be reduced in this figure, and throughout the preceding results, by increasing the driving frequency $\omega$ to further separate the driving-induced resonance from the background spin-mixing dynamics (up to the point at which the criterion for adiabaticity of the spatial dynamics 
does no longer hold).
\section{Discussion and outlook \label{sec:Discussion}}
In this work we have introduced a theoretical framework to understand the spin-mixing dynamics of a spinor BEC subject to a periodically modulated trapping potential. We have developed an effective few-state time-independent Hamiltonian, which is based on a separation of energy scales between spin and spatial dynamics, to describe the spin-mixing dynamics when the quadratic Zeeman energy $q$ is near resonant with the driving frequency of the trapping potential. From this model we systematically mapped out the mean-field dynamical phase diagram and the topology of the associated phase portrait. To assess the validity and robustness of our effective model, we employed numerical simulations based on the 3D time-dependent spinor GP equation, which describes the coupled dynamics of spin and spatial degrees of freedom in a spinor BEC, and studied a prototypical example of a spinor BEC confined in a periodically modulated harmonic trap. We found excellent agreement between the predictions of the GPE simulations and our effective model across a wide range of parameters, including clear signatures of critical spin-mixing dynamics near a series of dynamical phase transitions. The investigation of a harmonically trapped BEC complements prior experimental and theoretical work by ourselves and collaborators~\cite{Austin2024,Hardesty2023}, wherein we investigated spin-mixing dynamics of a spinor BEC subjected to a dynamical optical lattice. 

Our findings should be viewed in the light of related efforts focused on the study of Floquet-engineered spinor BEC dynamics via periodic driving of the quadratic Zeeman energy~\cite{Fujimoto2019floquetspinor,Evrard2019shapiro,Zhang2023spinorbit}.
In Ref.~\cite{Fujimoto2019floquetspinor}, Fujimoto and Uchino focus on a regime where the driving frequency $\omega$ of the time-dependent Zeeman energy $q(t) = q_{\mathrm{osc}}\cos(\omega t)$ dominates over all other energy scales, including the modulation amplitude $q_{\mathrm{osc}}$ and spin-independent interactions [see Eq.~(\ref{eqn:FullHam})].
The authors develop a leading order Floquet theory that leads to a description of spin-mixing with tunable relative contributions from spin-changing and spin-preserving collisions processes, equivalent to Eq.~(\ref{eqn:Heff}) of this work, although Fujimoto and Uchino retain the spatial degrees of freedom of the spinor BEC and investigate equilibrium physics. Of note, the requirement of a large driving frequency constrains the applicability of the theory to a perturbative regime close to the conventional undriven spinor BEC, i.e., $\eta = 2 -  (q_{\mathrm{osc}}/\hbar\omega)^2$
with $(q_{\mathrm{osc}}/\hbar\omega)^2 \ll 1$.
A similar setup is also investigated in Ref.~\cite{Zhang2023spinorbit} in the context of synthetic spin-orbit coupling. 
On the other hand, in Ref.~\cite{Evrard2019shapiro} Evrard and co-authors have reported an experimental and theoretical investigation of spin-mixing dynamics in the regime where the Zeeman energy is driven near-resonantly around a bias value $q_0$ such that $\hbar\omega \approx q_0$. By integrating out fast oscillating contributions, they developed secular equations of motion for the mean-field spin-mixing dynamics analogous to Eq.~(\ref{eqn:classicalEOM}) but with an effective $\eta = J_1(2q_{\mathrm{osc}}/\hbar\omega)$ and the constraint $q_{\mathrm{osc}} \leq q_0$. Therefore the tunability of the ratio $\eta$ is subject to a Bessel functional form of the driving parameters that imposes further constraints.
Lastly, we point out that Ref.~\cite{Li2019cavityspinor} has presented an alternative approach to Floquet-engineered dynamics mediated via the coupling of the internal Zeeman states of a spinor Bose gas to an optical cavity. Specifically, by pumping the cavity with a coherent laser field with periodically modulated intensity, one may obtain effective equations of motion describing parametric resonances in the spin-mixing dynamics equivalent to those in Ref.~\cite{Evrard2019shapiro}.

From a technical perspective, our work demonstrates an approach distint from Floquet-engineered spin-mixing dynamics between internal Zeeman states through time-dependent control of external degrees of freedom. In particular, our approach does not require fast control of time-dependent magnetic fields or strong off-resonant dressing fields, which can lead to undesirable heating and atom loss, to modulate the quadratic Zeeman energy. Moreover, we are able to control the ratio $\eta$ of the collision processes solely via the driving amplitude of the external potential as long as the driving remains adiabatic with respect to the spatial degrees of freedom (i.e., $\omega$ remains \textit{small} relative to the scale of the spin-independent interactions, as opposed to Ref.~\cite{Fujimoto2019floquetspinor}). Such a requirement can be independently controlled by modifying the external trapping of the condensate. For example, in a previous work by us and collaborators~\cite{Austin2024}, we loaded a BEC into an optical lattice on top of a background harmonic trap and periodically drove the lattice depth. In this case, the lattice set the energy scale of spatial excitations to be sufficiently large relative to the driving frequency, thus ensuring adiabaticity of the protocol.

More broadly, our approach provides an alternative means to control the sign and amplitude of the effective Zeeman shift. This can enable the observation of spin-mixing dynamics even at large absolute magnetic fields, and easy control of the initial spinor phase through the choice of driving scheme. 


Our results demonstrate that the sign of the effective Zeeman shift and the spinor phase can be manipulated by driving the nonlinear terms of the spinor Hamiltonian. The effective model operates in a regime where 
$q$ is close to resonance with the driving frequency, which is independently tunable. This tunability allows us to select the appropriate driving frequency to avoid resonant coupling to spatial excitations, thereby preserving the single-mode approximation.


The sensitivity near dynamical phase transitions in our system also highlights the potential for using spin dynamics to sense spatial excitations in spinor BECs, similar to spectroscopy in atomic and molecular physics. 
While in this work our results are all based on a mean-field description, which does not capture quantum correlations or entanglement, previous work by two of us has shown that metrologically useful entanglement can be generated near dynamical phase transitions \cite{Qingze2021}. The impact of Floquet-engineered spin-mixing on the dynamics of quantum fluctuation near the DPT will be explored in future work~\cite{SpinorQFI}.

\begin{acknowledgments}
Insightful discussions with Yingmei Liu and her experimental team and feedback on an earlier version of this manuscript are gratefully acknowledged. 
R.~J. L-S. acknowledges support by the National Science Foundation (NSF) through Grant No. PHY-2110052. This material is based upon work supported by the Air Force Office of Scientific Research under award number FA9550-24-1-0106. Q.G. acknowledges support by the NSF through Grant No. PHY-2409600 and support by Washington State University through the Claire May \& William Band Distinguished Professorship Award and the New Faculty Seed Grant. D.B. acknowledges support by  the NSF through Grant No. PHY-2409311. The computing for this project was performed at the OU Supercomputing Center for Education \& Research (OSCER) at the University of Oklahoma (OU) and the Kamiak High Performance Computing Cluster at Washington State University.
\end{acknowledgments}

\appendix

\section{Effective $1$D potentials\label{app:potentials}}
In the main text, we diagnose the dynamical phase diagram using an effective $1$D description of the system. Specifically, by using that the mean-field Hamiltonian $H_{\mathrm{mf},j}$ [Eq.~(\ref{eqn:Hmf})] defines a constant of motion we are able to reduce the pair of mean-field equations of motion for the conjugate variables $(\rho_0,\theta)$ [see Eq.~(\ref{eqn:classicalEOM})] to an equivalent one-dimensional problem. 

For the spinor phase, the dynamics are described by a single differential equation of the form, 
\begin{equation}
    \left(\frac{\hbar}{\mathcal{G}_0}\dot{\theta}\right)^2 + \frac{\mathcal{V}(\theta)}{\mathcal{G}_0} = 0 ,
\end{equation}
which is similar to a description of the motion of a classical particle traversing a potential well with position co-ordinate $\theta$. The function $\mathcal{V}(\theta)$ is given by
\begin{multline}
    \frac{\mathcal{V}(\theta)}{\mathcal{G}_0} = 16\tilde{E}_0 - 4(1+\tilde{q})^2 \\ - \eta\cos(\theta)\left[ 4 - 8\tilde{E}_0 + 4\tilde{q} + \eta\cos(\theta) \right],
\end{multline}
which is a function of the initial condition through the conserved quantity $\tilde{E}_0 = H_{\mathrm{mf},j}(\theta(0),\rho_0(0))/\mathcal{G}_0$ [computed from Eq.~(\ref{eqn:Hmf})].

For the fractional population, we specialize to the case of $\tilde{q} = 0$ as in the main text. We then obtain a differential equation of form 
\begin{equation}
    \left(\frac{\hbar}{\mathcal{G}_0}\dot{\rho}_0\right)^2 + \frac{\tilde{\mathcal{V}}(\rho_0)}{\mathcal{G}_0} = 0 ,
\end{equation}
where
\begin{equation}
   \frac{\tilde{\mathcal{V}}(\rho_0)}{\mathcal{G}_0} = 4\left[ \tilde{E}_0 + (\rho_0 - 1)(\tilde{q} + \rho_0) \right]^2 - \eta^2(\rho_0 - 1)^2\rho_0^2 .
\end{equation}



%

\end{document}